\documentstyle[12pt,psfig]{article}
\def\reference{\parskip 0pt\par\noindent\hangindent 0.5 truecm}

\begin{document}
%
%
\title{Angular Momentum Transfer in the Binary X-ray Pulsar GX 1+4}
%


\author{Greenhill J G $^{1}$ \and
 Galloway D K $^{2,3}$ \and
 Murray J R $^{4}$
} 

\date{}
\maketitle

{\center
$^1$ School of Maths and Physics, University of Tasmania, GPO Box 252-21, Hobart, Tasmania, 7005\\John.Greenhill@utas.edu.au\\[3mm]
$^2$ School of Maths and Physics, University of Tasmania, GPO Box 252-21, Hobart, Tasmania, 7005
$^3$ Research Centre for Theoretical Astrophysics, University of Sydney, Camperdown, NSW, 2006  \\Duncan.Galloway@utas.edu.au\\[3mm]
$^4$ Dept of Physics and Astronomy, University of Leicester, University  Rd, Leicester, LE1 7RH, UK\\jmu@star.le.ac.uk\\[3mm]
}

%
\begin{abstract}
We describe three presentations relating to the X-ray pulsar GX 1+4 at a 
workshop on magnetic fields and accretion at the Astrophysical Theory Centre, 
Australian National University on 1998, November 12-13. Optical and X-ray 
spectroscopy indicate that GX 1+4 is seen through a cloud of gravitationaly 
bound matter. We discuss an unstable negative feedback mechanism (originally proposed 
by Kotani et al, 1999) , based on 
X-ray heating of this matter which controls the accretion rate when the source is in a low X-ray luminosity state.  A deep minimum lasting $\sim 6$ hours 
occurred during observations with the RXTE satellite over 1996, July 19-21. 
The shape of the X-ray pulses changed remarkably from before to after the 
minimum. These changes may be related to the transition from neutron star 
spin-down to spin-up which occurred at about the same time. Smoothed particle 
hydrodynamic simulations of the effect of adding matter with opposite angular  momentum to an existing disc, show that it is possible for a number of 
concentric rings with alternating senses of rotation to co-exist in a disc. This could 
provide an explanation for the step-like changes in $\dot{P}$ which are observed in 
GX 1+4. Changes at the inner boundary of the disc occur at the same timescale as that 
imposed at the outer boundary. Reversals of material torque on the neutron star occur
at a minimum in $L_X$. 

\end{abstract}

Accretion, accretion disks --- (stars:) pulsars: individual (GX 1+4) --- stars:
winds, outflows --- X-rays: stars --- radiation mechanisms: thermal --- hydrodynamics

\bigskip

%
%

\section{Introduction}
The binary X-ray pulsar GX 1+4 is unique in several respects. It is the only 
known pulsar in a symbiotic system (V2116 Oph), $\dot{P}/P \sim2\%$ per year is 
the largest measured for any pulsar and the neutron star magnetic field is 
believed to be $\sim3\times10^{13}$ G - the strongest field in the known high
mass or low mass X-ray binaries.

The optical spectrum is that of an M giant plus a variable blue continuum and 
a forest of strong emission lines from H, HeI, FeII, [FeVII], [OIII] etc. The 
emission lines are believed to arise from photo-electric interactions of 
accretion disc UV photons in circumstellar matter (Davidsen, Malina \& Bowyer, 1977; 
Chakrabarty \& Roche, 1997). The blue continuum is generated by the disc.

The processes of angular momentum transfer in GX 1+4 are poorly understood. A 
long period of high luminosity and neutron star spin-up was followed by 
generally lower luminosity and spin-down interrupted by short episodes of 
spin-up. The average rates of spin-up and spin-down are similar but no clear 
correlation exists between $\dot{P}$ and X-ray luminosity $L_X$. The changes are not consistent with the standard accretion theory model (Ghosh \& Lamb, 1979).

In this workshop we investigated recent evidence relating to transitions
between spin-up and spin-down in GX 1+4. In chapter 2 we discuss evidence from X-ray and optical spectroscopy concerning the nature and distribution of the
circumstellar matter in this system. In chapter 3 we report sudden changes in 
X-ray pulse profiles associated with changes in $L_X$ and probably associated
with a brief transition to neutron star spin-up. In chapter 4 we compare the 
observed relation between $L_X$ and accretion torque with numerical 
simulations based on theoretical models of the accretion disc.
     
\section{Spectroscopic evidence on the nature of the neutron star environment}
GX 1+4 was observed with the ASCA X-ray satellite on 1994 September 14-15 and 
spectroscopy of the optical counterpart, V2116 Oph, was carried out at the 
Anglo-Australian Telescope using the RGO spectrograph on 1994 September 25-26. Photometry during 1994 August to October, using the Mt Canopus, Tasmania and 
Mt John, New Zealand 1 m telescopes showed little change in the source between the times of the X-ray and optical observations.

The X-ray spectroscopy showed considerable photo-electric absorption in the 
source region and strong iron line emisssion. The ionisation state of the iron 
(FeI-FeIV) shows that the $\xi$-parameter ($\equiv L_X/nr^2$, where n is the 
particle density of the circumstellar matter and r is the path length of the 
X-rays through this matter) $\leq 30 \, {\rm erg.cm.s^{-1}}$. Using the measured 
values of $L_X$ and $N_H (\sim nr)$ we estimate the characteristic scale of the attenuating matter distribution $r\geq 3\times10^{12}\,$ cm and $n\leq 7\times 
10^{10}\,{\rm cm^{-3}}$. The results of the optical spectroscopy were consistent with these conclusions. Using the Balmer line ratios and the calculations of Drake 
\& Ulrich (1980) we estimate the  
electron density $ n_e \sim 3\times 10^{10} - 10^{11}\,{\rm cm^{-3}}$ and plasma temperature $\sim 20,000K$ in the emission line region. The absence of FeIII and the presence of FeII lines supports this temperature estimate.

Using this information, Kotani et al (1999) propose a model in which the circumstellar matter is gravitationally bound to the neutron star during times of low $L_X$. This is consistent with the observed $H_\alpha$ line width ($\sim$ 2 AU) assuming doppler broadening from bound hydrogen at $r\sim 3\times 10^{12}$\, 
cm. The model provides an unstable negative feedback mechanism leading to 
large short term fluctuations in $L_X$ 
when the system is in a low intensity state. Increased accretion raises $L_X$ 
heating the trapped matter until the thermal velocity exceeds the escape 
velocity driving off trapped matter and suppressing accretion from the stellar wind. This occurs only at large distances from the neutron star so there is a 
delay with timescale $\sim$ the orbital period (several months) before accretion begins to decrease. Hence the accretion rate $ \dot M$ and $L_X$ will be unstable and  variable on timescales of months but relatively stable on longer timescales while the mean $L_X$ is low. If the ram pressure of the M giant wind (or
matter transferred by Roche Lobe overflow) becomes much higher than the 
thermal pressure in the trapped matter it will not be blown off by X-ray 
heating and the feedback mechanism will not be active. $L_X$ will be larger
and dependent only on the rate of mass flow from the M giant.
This mechanism requires very special conditions for its operation and is 
unlikely to apply to systems with supersonic winds as eg in Cen X-3 or 
Vela X-1.

\begin{figure}
 \begin{center}
 \psfig{file=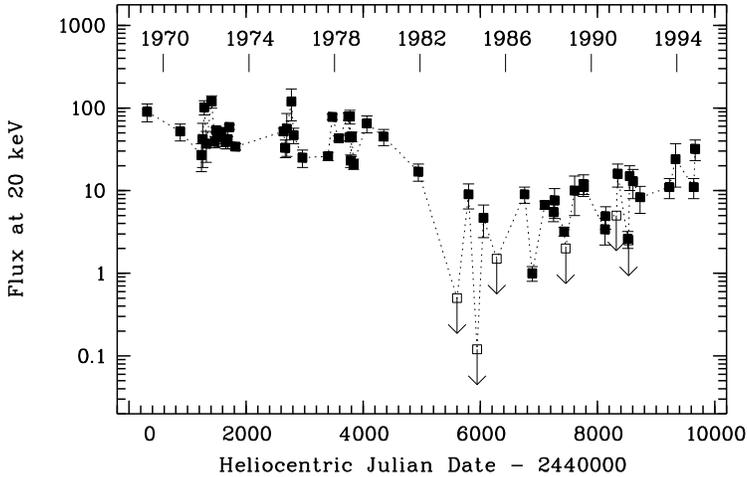,height=7cm}
 \caption{The time dependence of the 20 keV X-ray flux ($ \times 10^{-4} 
cm^{-2}s^{-1}keV^{-1 }$) from 1970 to 1994. The data are from a compilation by 
Greenhill \& Watson (unpublished report, 1994) of over 60  published and unpublished 
measurements. The filled squares represent positive detections and the open 
squares are $2 \sigma $ upper limits.}
 \label{fig1}            
 \end{center}
\end{figure}

The Kotani et al (1999) model provides a natural explanation for some aspects of the 
long term 
behaviour of GX 1+4. Greenhill \& Watson (unpublished report, 1994) collated 
the results of over 60 published measurements of GX 1+4 between 1971 and 1994.
Fig. \ref{fig1} is their estimate of the time dependence of the 20 keV X-ray 
flux during this period. Throughout the 1970's $L_X$ was large and relatively 
stable as expected when the feedback mechanism is not active. Subsequently 
the source was highly variable on timescales of order months and the mean 
value of $L_X$ was much 
lower. We suggest that the feedback mechanism was active during this period.
Another prediction is that large X-ray flares will be of shorter duration than
smaller flares. Large flares will blow off matter closer to the neutron star
and hence more rapidly affect accretion onto it. The pulsed flux history 
reported by Chakrabarty et al (1997) is qualitatively consistent with this 
prediction.

The model does not provide an explanation for the transition between intensity states. This may be caused by some long term instability in the giant 
companion. Nor does it  make any prediction concerning the direction of angular momentum transfer in this system. We note however that the negative feedback 
regime with a large diameter shell of low velocity trapped matter may be more 
conducive to the formation of a contra-rotating disc than the high luminosity 
regime when the ram pressure of the wind from the giant is higher and the wind extends much closer to the surface of the neutron star. 

\section{Spectral characteristics of GX 1+4 throughout a low flux episode}

GX~1+4 was observed using the Rossi X-ray Timing Explorer ({\it RXTE})
satellite (Giles et al. 1995) over 1996 July 19-21 during a period of
unusually low X-ray brightness for the source. For a detailed report see Galloway et al (1999) and Giles et al (1999). The countrate from the
Proportional Counter Array (PCA) aboard {\it RXTE} indicates that the mean
flux decreased smoothly from an initial level of $\approx 6\times
10^{36}\,{\rm erg\,s^{-1}}$ to a minimum of $\approx 4\times 10^{35}\,{\rm
erg\,s^{-1}}$ (20-60~keV, assuming a source distance of 10~kpc) before
partially recovering towards the initial level at the end of the
observation.

The pulse profiles (folded at the best-fit constant period) and the mean
photon spectra before and after the flux minimum show significant variation.
The observation is divided up into three distinct intervals based on the
mean flux. Interval 1 includes the start of the observation to just before
the flux minimum. Interval 2 spans $\sim 6 $ hours including the flux minimum, while during interval 3 the mean flux is rising steadily towards the end of 
the observation.

\begin{figure}
 \begin{center}
 \psfig{file=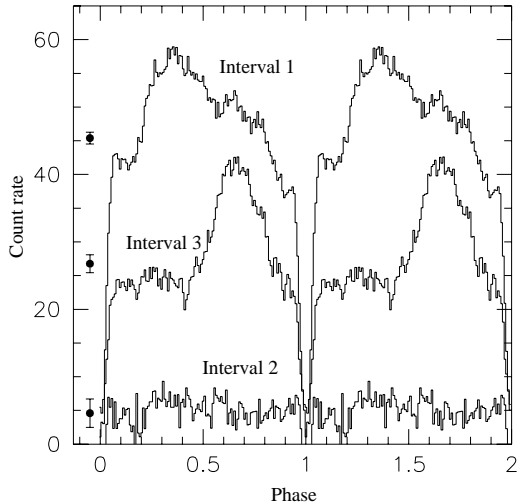,height=7cm}
 \caption{Pulse profiles for GX 1+4 during July 1996 folded on the best-fit
constant barycentre corrected period of $P=124.36568 \pm 0.00020$~s. The
data is taken from the PCA on board {\it RXTE} and spans the energy range
2-60~keV. Typical $1\sigma$ errorbars for each interval are shown at the left.}
 \label{fig2}            
 \end{center}
\end{figure}

The pulse profile is asymmetric and characterised by a narrow, deep
primary minimum (Fig. \ref{fig2}). During interval 1, the flux reaches a
maximum closely following the primary minimum; this is referred to as a
`leading-edge bright' profile. Pulsations all but cease during interval 2,
and in interval 3 the asymmetry is reversed, with the flux reaching a maximum
just {\it before} the primary minimum (`trailing-edge bright' profile). This
is the first observation of such dramatic pulse profile variations over
timescales of $< 1$~day.

Leading-edge bright profiles are generally associated with phases of
spin-down in GX~1+4, while trailing-edge bright profiles are mostly observed
during phases of spin-up (Greenhill, Galloway \& Storey, 1998).  Re-analysed data from the
regular monitoring of the source by the Burst and Transient Source
Experiment (BATSE) aboard the Compton Gamma-Ray Observatory ({\it CGRO})
indicate that the source switched from spin-down to spin-up $\approx 12$
days after the {\it RXTE} observation.  This suggests that the mechanism for
the pulse profile variations may be related to that causing the
poorly-understood spin period evolution in this source.

The best fitting spectral model (Galloway et al, 1999) is based on the work of Titarchuk (1994).
The principal component is generated by Comptonisation of a thermal input
spectrum at $\approx 1$~keV by hot ($kT \approx 8$~keV) plasma close to the
source with scattering optical depth $\tau_P \approx 3$. Additional
components include a gaussian to fit Fe K$\alpha$ line emission from the
source and a multiplicative component representing photoelectric absorption
by cold material in the line-of-sight.  Variations in the mean spectrum over
the course of the observation are associated with a dramatic increase in the
column density $n_H$ from $13\times 10^{22}$ to $28\times 10^{22}\,{\rm
cm}^{-2}$ between intervals 1 and 3, and also with significant
energy-independent variations in the flux.

Similar spectral variations were seen in 4U 1626-67 before and after the spin
rate transition in that source (Yi \& Vishniac, 1999). This strengthens the 
argument that the pulse profile and spectral changes reported here were 
associated with the torque reversal in GX1+4 reported by Giles et al. (1999).

Pulse-phase spectral fitting indicates that variations in flux with phase
can be accounted for by changes in the Comptonised model component, with in 
particular variations in the fitted optical depth $\tau_P$ and the component
normalisation $A_{\rm C}$ accounting for the phase dependence. The spectral
fits suggest that the soft input photons originate from the neutron star
poles, and are subsequently Comptonised by matter in the accretion columns.
The sharp dip in the pulse profiles is then tentatively identified with the
closest passage of one of the magnetic axes to the line of sight. More
details of the spectral analysis can be found in Galloway et al. (1999).

\section{SPH modelling of a counter rotating disc}

The new continuous monitoring data obtained by BATSE 
(Bildsten et al. 1997, Nelson et al. 1997, Chakrabarty et al. 1997)
have yielded values for the mass accretion rate (X-ray
luminosity) and the accretion torque (change in spin frequency) 
on a regular basis for a large number of X-ray pulsars.
This data set has allowed a detailed comparison of the observed relation
between accretion rate and torque, and that predicted by theoretical
models.

The observations are not consistent with the standard Ghosh \&
Lamb (1979) model, since this predicts a clear correlation between spin-up
and an increase in X-ray luminosity, whereas the observations show a
variety of behaviour, with spin-up or spin-down occurring at the same
apparent luminosity.  

Numerical simulations (see Ruffert, 1997 and references therein) have shown
that it is possible to form temporary accretion discs with alternating
senses of rotation in wind-accreting systems.  Nelson et al. (1997) made the
{\em ad hoc} suggestion that many observational features of some systems
that are normally thought to contain discs (GX~1+4, 4U~1626-67) would be
explained if they were accreting from discs with alternately prograde and
retrograde senses of rotation.  Previously, Makishima et al. (1988), Dotani
et al. (1989) and Greenhill et al. (1993) had also sought to explain the
rapid spin-down of GX~1+4 in terms of accretion from a retrograde disc.

If the secondary star is feeding the
accretion disc via Roche lobe overflow, as is almost certainly the
case in 4U~1626-67, it is hard to conceive how a
retrograde disc could ever come about. However, in the case of GX~1+4,
the suggestion is not unreasonable. This X-ray pulsar is unique in the
sense that it is accreting from a red giant or AGB star wind (Chakrabarty
\& Roche 1997), and is in a very wide orbit. Estimating the timescale
of disc reversal for accretion from such a wind, one obtains a
timescale of the order of years, and the disc would form at a large radius
($\sim 10^{13}$ cm) so that the inner part of the accretion flow is
expected to be like a normal accretion disc. A timescale of years
corresponds well with the timescale on which the accretion
behaviour in GX~1+4 is observed to change, with a negative correlation
between accretion rate and spin-up in some phases while the disc would
be retrograde, and a positive one at other times when it is prograde
(Chakrabarty et al. 1997). Thus, this system is ideally suited to
study the possibility of forming retrograde discs, since the timescale
for disc reversal would be much longer than that of 
the torque fluctuations on a timescale of one day or less that are
common in all types of X-ray pulsars. 
In the systems that accrete from a
fast wind, the two timescales are comparable, and the effects will
be difficult to separate.

Two dimensional 
smoothed particle hydrodynamics (SPH) simulations were used to investigate
the interactions of an existing accretion disc with material coming in
with opposite angular momentum. See Murray, de~Kool \& Li, 1999 for
more details of the calculations. 
Ideally we should like to simulate the entire accretion
disc. However, for GX~1+4, this would require resolution over several
decades in radius. Instead we completed two separate simulations: the
first being of the inner, viscously dominated region which for GX~1+4 we expect
to extend from the neutron star magnetosphere out to a radius $r
\simeq 5 \times 10^{10}$ cm; and the second being 
of the outer disc in which the dynamical mixing of material with
opposite angular momentum dominates. 

\begin{figure}
\begin{minipage}[t]{75mm}
\psfig{file=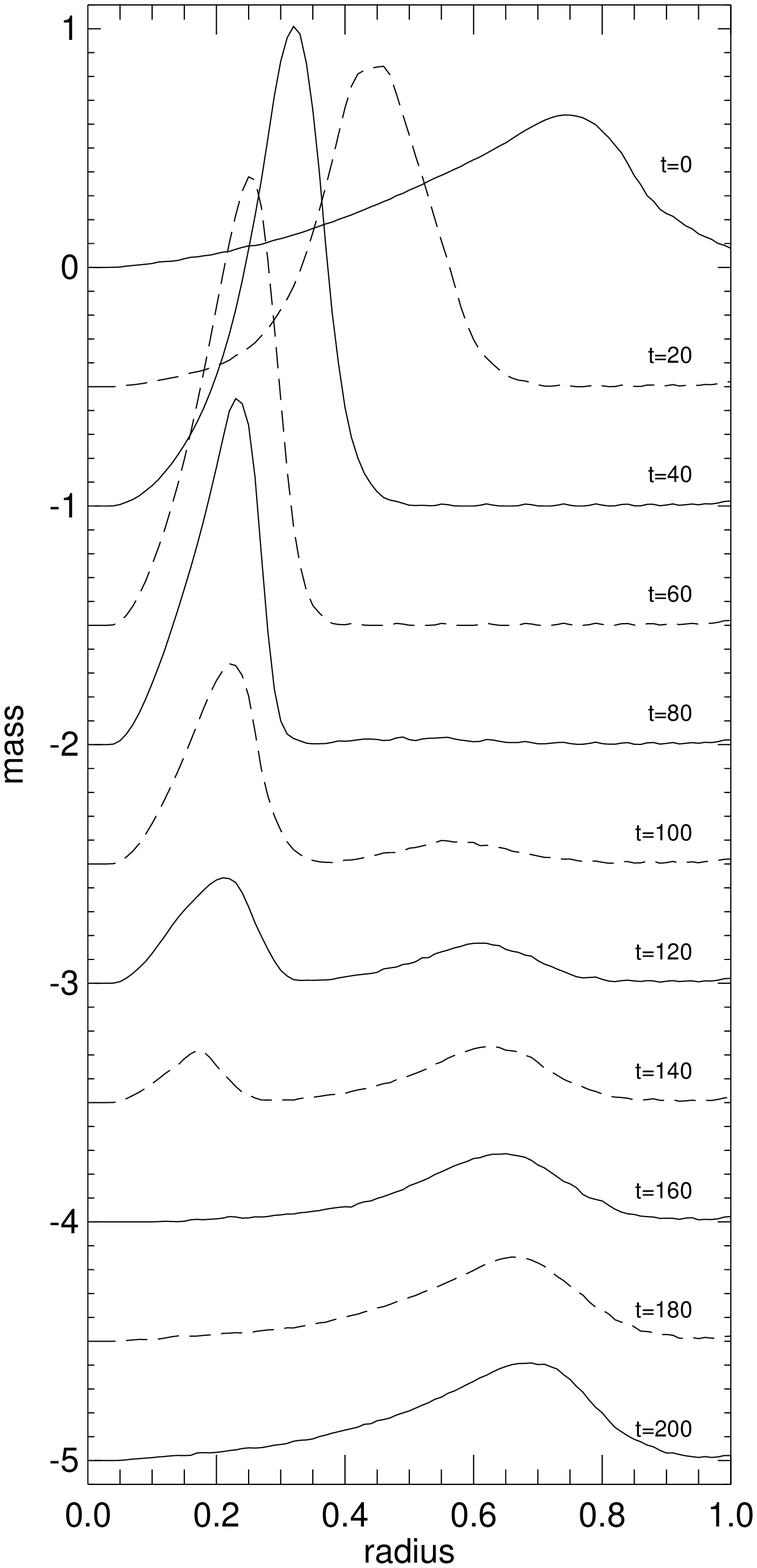,width=7.5cm}
\caption{The evolution of the radial mass profile 
of a viscous ring that is subjected at its outer
edge to the addition of material with opposite specific angular
momentum.}
\label{fig:sim1}
\end{minipage}
\hspace{\fill}
\begin{minipage}[t]{75mm}
\psfig{file=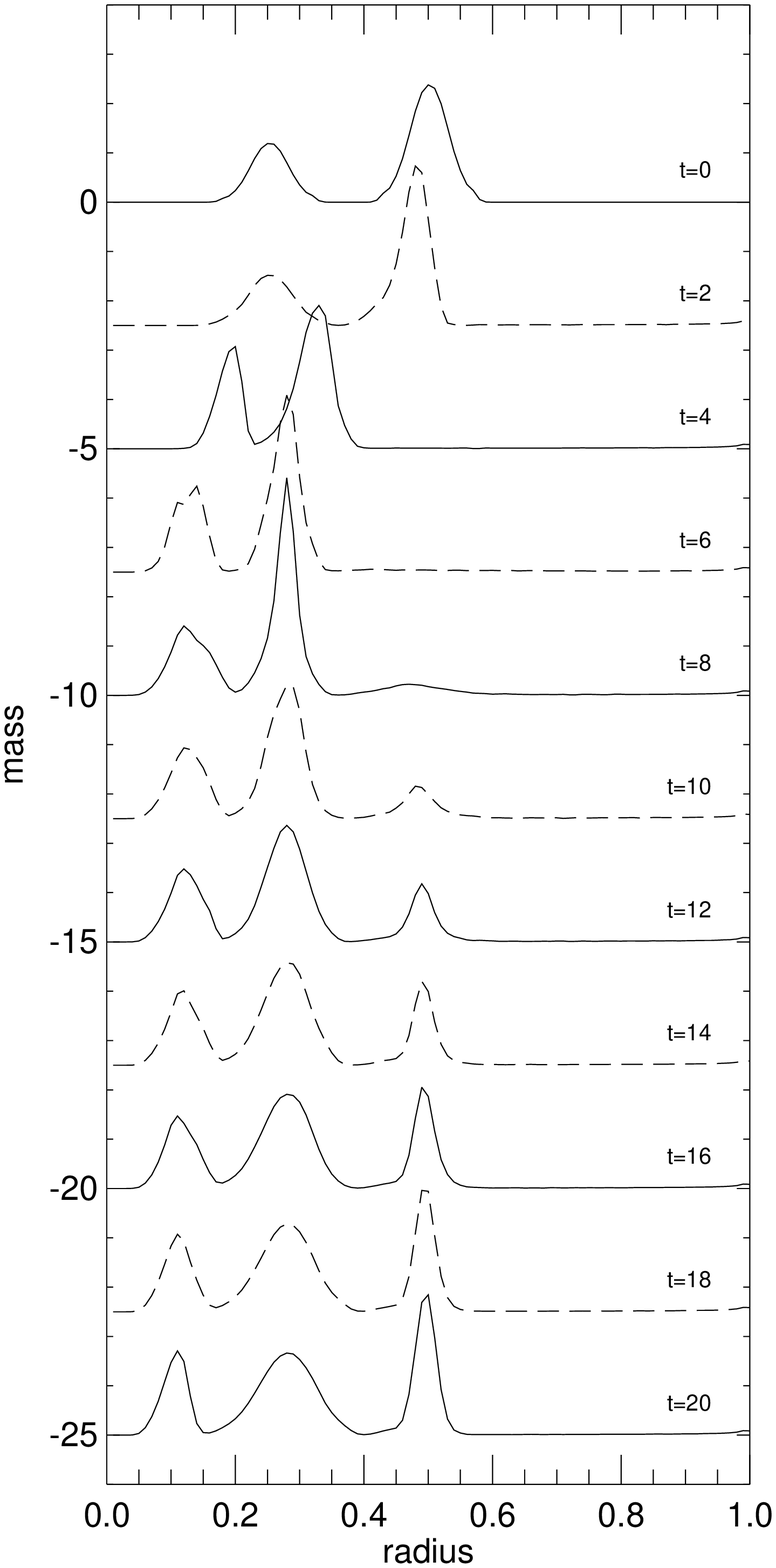,width=7.5cm}
\caption{Evolution of two concentric counter-rotating rings that are
subject to mass addition (rotating in the sense of the inner ring)}
\label{fig:sim2}
\end{minipage}
\end{figure}

\begin{figure}[htb]
\psfig{file=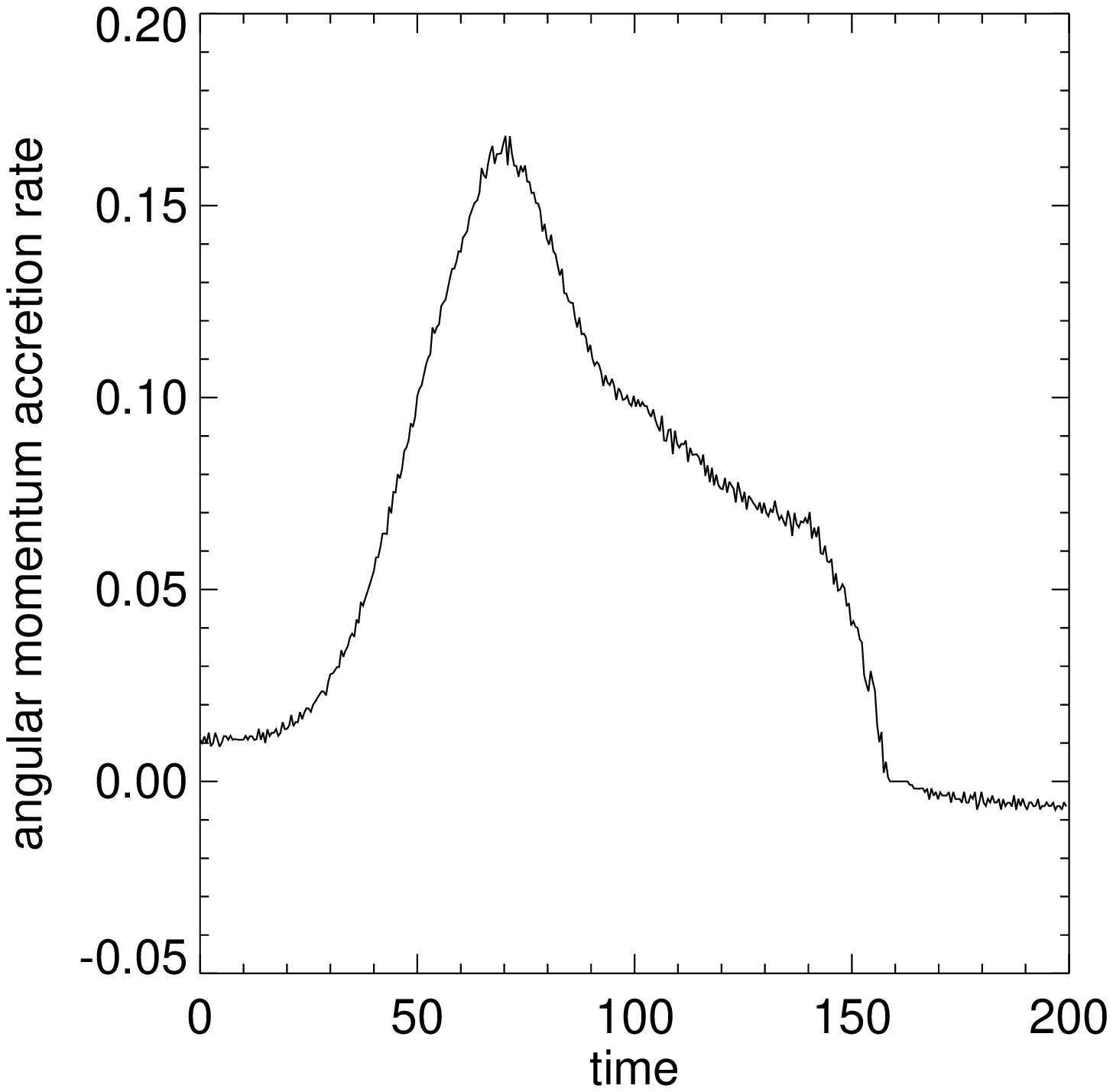,width=7.5cm}
\caption{Angular momentum accretion rate (material torque) for the
simulation of the viscous inner disc that is illustrated in 
figure~\ref{fig:sim1}.}
\label{fig:torque}
\end{figure}

We found that in the inner disc, once the  sense of 
rotation of inflowing material was reversed (figure~\ref{fig:sim1}), the 
existing disc was rapidly driven inside the circularisation radius of the new
counter-rotating matter. Further evolution  occurred on the
viscous time scale, with the initial disc slowly being accreted at the
same time as a second counter-rotating disc formed outside it. We
found that the rate of angular-momentum accretion (i.e. the material
torque, shown in figure~\ref{fig:torque}) was proportional to the mass accretion rate. The material torque
did not change sign until the initial disc had been entirely
consumed. The change in sign of the torque was accompanied by a {\bf
minimum} in the accretion luminosity.

The second calculation 
(figure~\ref{fig:sim2})
began with two
counter-rotating rings with Gaussian density profiles. New material,
with the same sense of rotation as the inner ring, was then added at a
constant rate at the outer boundary. As with the first calculation we
found the initial rings were rapidly driven in until they lay within
the circularisation radius of the newly added  material. We had
anticipated a catastrophic cancellation of angular momentum followed
by radial inflow once the rings interacted. This did not
happen. Instead the rings remained cohesive with a well defined gap
between them. The newly added material then formed a third, outer ring. 
We concluded that, if the external mass
reversal timescale is significantly shorter than the viscous timescale
at the circularisation radius, a number of concentric rings with
alternating senses of rotation could be present between the
circularisation radius and the radius at which the viscous timescale
is comparable to the reversal timescale. 
These simulations neglected three-dimensional effects, and didn't
account for the unstable wind feeding non-coplanar material onto the
system. However we are confident that further work will not alter our
main conclusion that changes at
the inner boundary of the disc occur on the same timescale as that
imposed at the outer boundary. Furthermore we find that material torque
reversals occurring as a result of a disc reversal, would do so during
an accretion luminosity minimum.

\section{Discussion \& Conclusions}
Optical and X-ray spectroscopy indicate that a circumstellar cloud or thick 
disc extends to at least $3 \times 10^{12}$ cm from the neutron star. This has a mean density $\sim 7\times 10^{10}\,{\rm cm^{-3}}$ and temperature $\sim 20,000$ K 
in the emission line region. We describe a model, proposed by Kotani et al 
(1999), in which, during the low intensity state, accretion is controlled by 
X-ray heating leading to an unstable  negative feedback mechanism. This 
maintains $L_X$ at relatively low levels but highly variable on timescales of 
$\sim$ months. Physical conditions in the trapped matter region may be more 
conducive to formation of a contra-rotating disc and neutron star spin-down 
when this feedback mechanism is active. When the accretion rate is high, the 
mechanism fails and $L_X$ is higher and more stable. The transition between 
states is presumably controlled by some instability in the giant companion.

The X-ray pulse profiles from GX 1+4 changed remarkably during an observation 
by the RXTE satellite over 1996 July 19-21. The profiles were asymmetric and
'leading edge bright' during the early part of the observations when $L_X$
(20-60 keV) was $\sim 6 \times 10^{36}\,{\rm erg s^{-1}}$ (source distance 10 kpc).
After an interval of $\sim 6$ hr. when $L_X$ was $\sim 10$ times lower, the 
intensity increased towards the initial level but the profiles had changed to 
'trailing edge bright'. The change in profile may be related to a transition 
from spin-down to spin-up which was detected by the BATSE experiment on CGRO
at about the same time. According to Greenhill et al (1998) leading edge 
bright/trailing edge bright profiles are normally associated with neutron star 
spin-down/spin-up repectively.

The X-ray spectrum during the RXTE observations was best characterised by
Comptonised thermal emission with iron line emission and photo-electric
absorption by cold matter in the source region. The column density $n_H$
doubled between the the early and late phases of the observation and showed
significant variation on timescales as short as 2~h.  This change is too
short to be associated with the feedback mechanism discussed above. The
extra absorbing matter in the line of sight must be situated much closer to
the neutron star.

Two dimensional SPH simulations have been used to investigate the
interactions of an existing accretion disc with incoming matter having
opposite angular momentum. The simulations showed that a counter-rotating
disc was formed outside the existing disc which quickly shrunk inside the
circularisation radius of the outer disc. The inner disc was accreted on the
viscous timescale. The torque did not change until this disc was fully
consumed and torque reversal was accompanied by a minimum in $L_X$. If the
external mass reversal timescale is significantly shorter than the viscous
timescale at the circularisation radius, a number of concentric rings with
alternating senses of rotation can co-exist. Changes at the inner boundary
of the disc occur at the same timescale as that imposed at the outer
boundary. Material torque reversals occur at a minimum in $L_X$. 

The net torque on the neutron star depends also on magnetic torques due to 
linkage with disc matter both inside and outside the co-rotation radius 
(Ghosh \& Lamb, 1979, Li \& Wickramasinghe, 1997). The two transitions to 
spin-up reported by Chakrabarty et al (1997) occurred 
when $L_X$ was increasing by more than an order of magnitude from very low 
levels. Conversely the transition to spin-down was associated with a similar 
magnitude \it decrease \rm to a very low level. Such transitions could be 
caused by a disc having alternate zones with prograde and retrograde motion.
The BATSE record (Chakrabarty et al, 1997) shows that, during intervals of 
monotonic spin-up or monotonic spin-down, GX 1+4 and several other wind fed  
sources make step like transitions from one value of $\dot{P}$ to another. 
Hence, if the disc velocity profile has abrupt changes 
switching the sense of rotation between different zones, as 
discussed in section 4, step like changes in the magnitude of $\dot{P}$ will
occur as the matter is transported inwards. The analysis by Wang \& Welter
(1981) indicates that asymmetry in pulse profiles 
may be a consequence of an asymmetry in the accretion flow onto the polar cap 
region. Hence, the reversal in the asymmetry of the pulse 
profiles observed in the RXTE data could be a consequence of 
accretion flow changes as the direction of rotation of the inner edge of the 
disc reversed. This occurred at a minimum in $L_X$ as predicted by the SPH 
modelling.




\section*{Acknowledgements}
We acknowledge Martijn de Kool and Jianke Li for substantial contributions to 
the work described in section 4.  Barry Giles did much of the work on the XTE 
X-ray data. Michelle Storey and Kinwah Wu helped organise the workshop 
sessions on GX 1+4. We are grateful to the Astrophysical Theory Centre and to 
Martijn (again) for organising a very successful workshop.


\section*{References}






\reference Bildsten, L., et al. 1997, ApJS, 113, 367

\reference Chakrabarty, D., Bildsten, L., Grunsfeld, J.M., Koh, D.T., Nelson, R.W., Prince, T.A. \& Vaughan, B.A.. 1997, ApJ, 481, L101

\reference Chakrabarty, D. \& Roche, P. 1997, ApJ, 489, 254

\reference Davidsen, A., Malina, R. \& Bowyer, S. 1977, ApJ, 211, 866

\reference Dotani, T., Kii, T., Nagase, F., Makishima, K., Ohashi, T., Sahao,
T., Koyama, K.\& Tuohy, I.R. 1989, PASJ, 41, 427

\reference Drake, S.A.\& Ulrich, R.K. 1980, ApJS, 42, 351

\reference Galloway, D.K., Giles, A.B., Greenhill, J.G. \& Storey, M.C., 1999,
  submitted to MNRAS

\reference Ghosh, P.\& Lamb, F.K. 1979, ApJ, 234, 296

\reference Giles, A.B., Jahoda, K., Swank, J.H.\& Zhang W. 1995, PASA, 12, 219

\reference Giles, A.B., Galloway, D.K., Greenhill J.G., Storey M.C. \& Wilson, C.A. 1999,
  submitted to ApJ

\reference Greenhill, J.G., et al. 1993, MNRAS, 260, 21

\reference Greenhill, J.G., Galloway, D.K. \& Storey M.C. 1998, PASA, 15, 2, 254

\reference Kotani, T., Dotani, T., Nagase, F., Greenhill, J., Pravdo, S.H. \&
Angelini, L. 1999, ApJ, 510, 369

\reference Li, J. \& Wickramasinghe, D.T. 1997, MNRAS, 286, L25

\reference Makishima, K., et al. 1988, Nature, 333, 746

\reference Murray, J.R., De~Kool, M. \& Li, J. 1999, ApJ, 515, 738 

\reference Nelson, R.W., et al. 1997, ApJ, 488, L117

\reference Ruffert, M. 1997, A\&A, 317, 793

\reference Titarchuk, L. 1994, ApJ, 434, 570

\reference Wang, Y.M. \& Welter, G.L. 1981, A\&A, 102, 97 

\reference Yi, I. \& Vishniac, E. 1999, ApJ, 516, L87

\end{document}